\documentclass[aip,reprint,superscriptaddress]{revtex4-2}
\usepackage{amsmath}
\usepackage{amssymb}
\usepackage{graphicx} 
\usepackage{dcolumn} 
\usepackage{bm} 
\usepackage{times}
\usepackage{hyperref}
\usepackage{soul}
\usepackage[english]{babel}
\usepackage[latin1]{inputenc}   
\usepackage{color,soul}
\usepackage{hyperref}
\usepackage{graphicx} 
\hypersetup
{
	bookmarksopen=true,
	pdftitle="",
	pdfauthor="",  
	pdfsubject="",  
	pdftoolbar=true,  
	pdfmenubar=true,  
	pdfhighlight=/O,  
	colorlinks=true,  
	pdfpagemode=UseNone,  
	pdfpagelayout=SinglePage,  
	pdffitwindow=true,  
	citecolor=blue,
	linkcolor=blue,
	urlcolor=blue  
}

\usepackage{wasysym}
 
\begin{document}
\title{Data analysis methods for powder x-ray diffraction intensity under laser-driven dynamic compression at Omega and NIF laser facilities}
\author{Marius Millot}
 \email{millot1@llnl.gov}
\author{Federica Coppari}
\author{ Amy Lazicki}
\author{Jon H. Eggert}
\affiliation{Lawrence Livermore National Laboratory, Livermore, CA 94550, USA}
\begin{abstract}
Powder x-ray diffraction (PXRD) under laser-driven dynamic compression is a powerful tool to investigate material response to extreme pressure, temperature and strain rates. Robust PXRD platforms have been developed at kJ and MJ laser facilities worldwide including the Powder X-Ray Diffraction Image Plate (PXRDIP) at the Omega Laser Facility at the Laboratory for Laser Energetics (LLE) and the TARget Diffraction In Situ (TARDIS) at the National Ignition Facility (NIF). Here we present further developments of data analysis methods focused towards improving the fidelity of the PXRD intensity determination for these  platforms. We illustrate these methods by discussing how they can be implemented in a data analysis package and applied to shock compression data on diamond near 1 TPa. We discuss using the XRD signal from the collimating pinhole or a layer of un-compressed material in the sample package as \textit{ in-situ} references for XRD intensity. We detail how to compare data collected with different x-ray sources and how to account for thermal damping of XRD signal when comparing XRD from a shock-compressed, hot material with the reference material at ambient.
\end{abstract}
 
\maketitle
\section{Introduction }
 The Powder X-Ray Diffraction Image Plate (PXRDIP) platform\cite{rygg2012} was initially developed at the Omega 60 Laser Facility at the Laboratory for Laser Energetics (LLE), in Rochester, NY. It was later adapted to the LLE Omega EP Laser Facility and has now been fielded in more than 250 shot-days combining Omega 60 and EP. The TARget Diffraction In Situ\cite{rygg2020,rygg2026} (TARDIS) at the National Ignition Facility (NIF) in Livermore, CA has been used in more than 200 NIF experiments.  Important results obtained with the PXRDIP platform include the discovery of superionic water ice XVIII\cite{millot2019a}, of the B2 phase of MgO\cite{coppari2013} and studies on the melting line of Ta\cite{kraus2021} and MgO\cite{wicks2024}. TARDIS highlights include documenting the melting on iron near 1 TPa\cite{kraus2022} and the persistence of the diamond structure of carbon under ramp compression to 2 TPa\cite{lazicki2021}. 
 \begin{figure}[!htb]
  
 	\centerline{\includegraphics[width=.4 \textwidth,trim=40 80 450 90,clip]{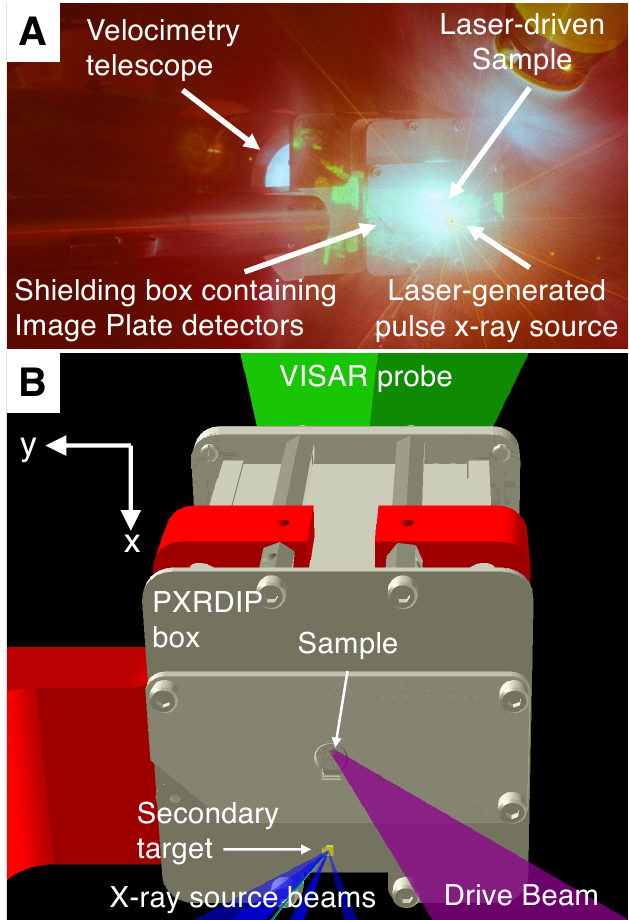}}     
 	\caption{\textbf{X-ray diffraction experimental configuration with the PXRDIP diagnostic at the Omega EP laser facility.} (\textbf{A}) Time-integrated photograph of a steady shock experiment. (\textbf{B}) Simplified 3D model. Here, one beam (purple) drives the sample package while up to three beams (blue) are tightly focused onto the secondary target (24 mm away, $\sim$ 23 degrees from normal incidence) to produce an x-ray flash. Image plate detectors line the interior walls of the $\rm 50 mm \times 50 mm \times 75 mm$  PXRDIP shielding box to record the x-ray diffraction pattern. An aperture in the back of the box provides a line of sight for the VISAR velocimetry (VISAR)  and pyrometry (SOP) measurements. The standard PXRDIP platform configuration sets the 1D VISAR imaging direction along the x axis \textit{i.e.} along the vertical axis of the target chamber. This means that VISAR documents the drive planarity along that direction, which corresponds to XRD data on the Up and Down panels (see Section \ref{sec:Projection into 2TT-radius}). The \textit{z} axis is along the VISAR line of sight, oriented towards the VISAR telescope. }\label{fig:Visrad1}	
 \end{figure}  
Although focusing about 10 kJ of ultra-violet laser light down to $\sim$1 millimeter diameter enables spatially-uniform, well controlled dynamic compression on samples approximately 100 microns thick, conducting \textit{in-situ} x-ray diffraction (XRD) measurements remains highly challenging. Among the many challenges to be addressed, one is to accurately determine the diffraction geometry, in other words the relative position of the source, sample, reference materials and each of the image plate detectors and account for the non-standard orientation of the image plate detectors with respect to the diffracted x-rays. In addition, PXRDIP raw data are usually dominated by strong, spatially variable background generated by the 10-50 ns long, high-intensity ($10^{12-14}W/cm^2$) laser ablation drive used to achieve precisely controlled shock or ramp compression. Other difficulties are associated with laser-generated plasma x-ray sources: the significant bandwidth of the plasma line emission contributes to peak broadening, while weaker continuum emission may be bright enough to generate intense Laue diffraction spots if the sample package includes single crystals. Due to these challenges, most studies to date with the PXRDIP and TARDIS platforms focused on documenting the presence of Bragg peaks and determining the diffraction angles to infer the density.

Here we build on the PXRDIP data analysis methods\cite{rygg2012} as well as refinements developed for TARDIS\cite{rygg2020} to improve the fidelity of the PXRD intensity determination. Better constraining XRD intensity determination from PXRDIP and TARDIS experimental data offer promising opportunities to study phase transformation and phase coexistence\cite{millot2025} and may enable identifying subtle phase changes such as isostructural transition to superionic behavior\cite{devilla2025}. While several shock studies with PXRDIP or TARDIS have been reported \cite{wang2015b,kraus2021,wicks2024,poole2024,marshall2022} the previous descriptions of the PXRDIP and TARDIS platforms \cite{rygg2012,rygg2020} focused the discussion on ramp compression experiments. Here we illustrate these methods with an experiment on shock compressed diamond near 1 TPa collected at the Omega EP laser facility as an example and by describing in detail their implementation in our \textsc{AnalyzePXRDIP} data analysis package implemented in \textsc{IgorPro}. We note that although the discussion of the analysis methods focus on the PXRDIP experimental platform available at the Omega laser facility, methods are also applicable to the TARDIS diagnostic available at the NIF and should be easily transferable to XRD experiments at the Laser Fusion Research Center (LFRC)  \cite{sun2024} .

 This article is organized as follows. We describe the experimental configuration for Powder X-ray diffraction at Omega EP in Sec.\ref{Sec:Experimental configuration for PXRD at Omega EP}, and an example dataset on shock compression of diamond near 700 GPa in Sec.\ref{Sec:Compression}. Sec.\ref{sec:PXRDIP data analysis workflow} describes the PXRDIP data analysis workflow, beginning with a discussion of the geometry and raw data (Sec.\ref{Sec:Geometry and data format}) and including a discussion of projections of the data into diffraction versus azimuth angle (Sec.\ref{sec:Projection into 2TT-phi}) and into diffraction angle versus scattering radius (Sec.\ref{sec:Projection into 2TT-radius}). We then discuss XRD intensity corrections applied before background subtraction (Sec.\ref{sec:XRD Intensity corrections before background subtraction}), SNIP background subtraction and further intensity corrections (Sec. \ref{sec:SNIP background subtraction and further intensity corrections}). Next we describe our post-processing and x-ray diffraction intensity analysis
  (Sec.\ref{sec:Post-processing and  x-ray diffraction intensity analysis}) including a discussion on intensity lineouts and fits (Sec.\ref{Sec:Intensity lineouts and multi-peak fits}) and XRD intensity scaling by the compressed sample volume (Sec.\ref{sec:volumescaling}), by the x-ray source brightness(Sec.\ref{sec:scaling of the XRD intensity to the expected number of x-ray photons}) and (Sec.\ref{sec:Evolution of the XRD intensity relative the ambient platinum signal}). Finally, we discuss the possibility of using an uncompressed reference layer such as the unshocked layer in a shock compression experiment (Sec.\ref{sec:Evolution of the XRD intensity relative the unshocked diamond signal}). Note that in this last section we restrict the discussion to the simple case of a monoatomic cubic crystal such as diamond, with an initially polycrystalline material.

   \section{Experimental configuration for Powder X-ray diffraction at Omega EP}\label{Sec:Experimental configuration for PXRD at Omega EP}
  
  \begin{figure*}[!ht]
 
  	  	\centerline{\includegraphics[width=.8 \textwidth,trim=40 250 200 130,clip]{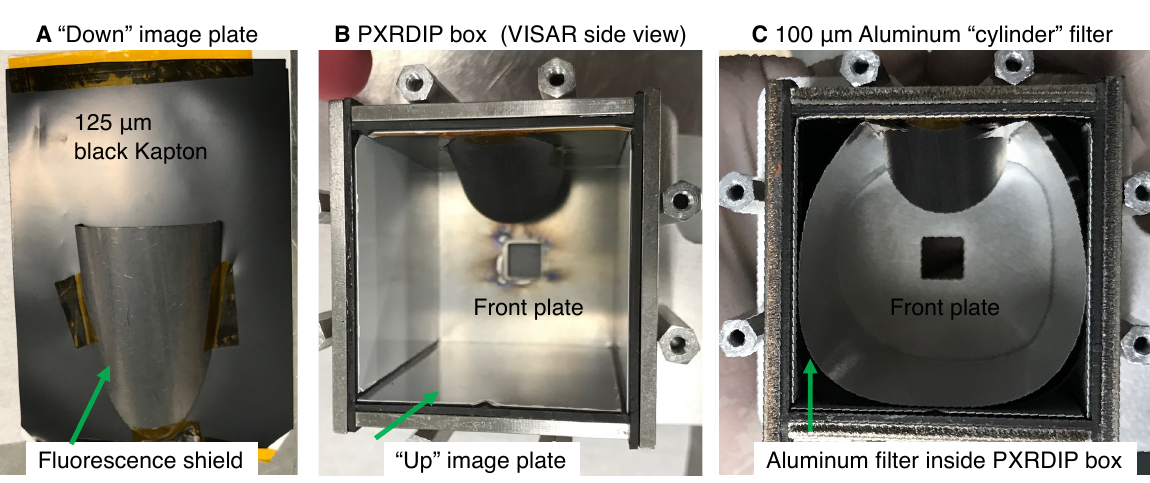}}     
  	\caption{\textbf{PXRDIP fluorescence shield and x-ray filters.} (\textbf{A}) Photograph of the \textit{Down} image plate (IP) before insertion into the PXRDIP box, revealing the presence of a flat, 125 $\mu m$ thick black Kapton filter covering the IP, and a Ta shield positioned to block x-ray fluorescence induced in the IP's phosphor layer by the intense direct x-ray beam impinging onto the {Down} IP. (\textbf{B}) Photograph of the PXRDIP box, after installation of the four main IPs, viewed from the VISAR side. (\textbf{C}) For the example experiment discussed here, A 100 $\mu m$ thick rectangular aluminum filter is then shaped into a {cylinder} and inserted inside a similar {cylinder} made of 125 $\mu m$ thick black Kapton.}\label{fig:PXRDIP_FIlters}	
  \end{figure*}  
     
 We briefly describe the PXRDIP platform at the Omega EP laser to provide context for the data analysis discussion. Further details on the experimental configuration can be found in Ref.\cite{rygg2012}. Because the NIF TARDIS platform\cite{rygg2020,rygg2026} is closely related to the PXRDIP, the methodology described in this work can be readily implemented to process TARDIS data. A planar sample package is placed on the front plate of a PXRDIP metallic box (see Fig. \ref{fig:Visrad1}). Omega EP is used in its \textit{Long Pulse} configuration, with four individually controlled laser beams delivering several kJ of 351 nm light in precisely temporally shaped pulses\cite{guardalben2020}, at a 23.2 degree incidence. Omega EP experiments typically use one or two beams to drive the sample with up to  $\sim$ 10 ns long pulse and total energy up to $\sim$ 4500 J/beam and EP SG8-1100 super-gaussian distributed phase plates (DPPs). The laser intensity in the best focus plane of these DPPs can be accurately modeled as $I(r) = I_0~  exp(-(\frac{r}{r_0})^\eta)$ where $I(r)$ is the laser intensity as a function of the distance from the beam center $r$, $I_0$ is the peak intensity, $r_0=483 \mu m$ is the characteristic radius, and $\eta=8.57$ is the power of the super-Gaussian. Using a scaling law\cite{fratanduono2011b, drake2006} for direct-drive laser ablation  $P_{Abl}\propto I^{2/3}$ suggests  that the diameter over which the ablation pressure is greater than 95\% of the maximum is about 720 $\mu m$.
 
  Two or three additional laser beams can be tightly focused (down to $\sim$ 200 $\mu m$ beam size, without phase plates) onto a secondary target to produce an x-ray flash for the x-ray diffraction measurement\cite{coppari2019}. Each beam delivers $\sim$ 1250 J in a super-gaussian 1 ns flat-top pulse -- an irradiance of $\sim 3\times 10^{15} \rm W/cm^2$ -- with a co-timing accuracy better than 100 ps. This produces strong pulses of quasi-monochromatic x-ray emission dominated by the He-$\alpha$ complex  with a 40 eV bandwidth near 10.249 keV  (1.2097 $\rm \mathring{A}$) for Ge while Cu sources have a 89 eV bandwidth near 8.3680 keV (1.4816 $\rm \mathring{A}$). A flat crystal x-ray spectrometer\cite{thorn2018} is usually fielded to document the x-ray emission. Details on laser-driven x-ray sources for XRD are discussed in Ref.\cite{coppari2019} (Omega 60 and EP) and in Ref.\cite{krygier2022a} (NIF).
  
  A full suite of laser diagnostics\cite{guardalben2020} records the precise pulse shape and relative timing of the various laser beams. This enables determining the timing of the x-ray flash with respect to the drive laser. For each shot a two-channel line-imaging Doppler Velocimetry Interferometer System for Any Reflector (VISAR) \cite{celliers2023} (operating at 532 nm at the Omega Laser Facility) and a line-imaging Streaked Optical Pyrometer (SOP) \cite{gregor2016}(collecting visible light with a wavelength $\lambda >590 nm$ at the Omega Laser Facility) diagnose the compression history and help determine the thermodynamic state characterized with XRD.  Details on the drive characterization for ramp compression XRD experiments at Omega are discussed in Ref.\cite{coppari2022}.

  \section{Example dataset: shock compression of diamond near 700 GPa}\label{Sec:Compression}
  \begin{figure}[!b]
  	  	\centerline{\includegraphics[width=.48\textwidth,trim=10 140 330 90,clip]{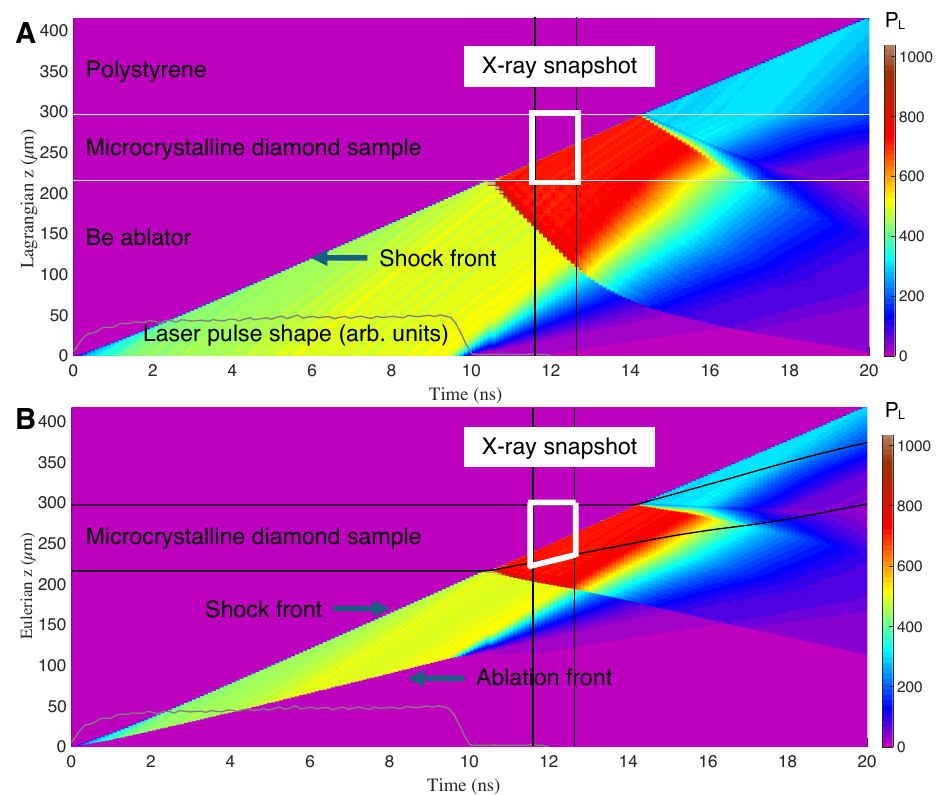}}     
  	\caption{\textbf{1D hydrodynamic simulation of a steady shock XRD experiment}. $z-t$ Longitudinal stress ($P_L$) colormap (in GPa) as a function of time ($t$) and Lagrangian (\textbf{A}) or Eulerian (\textbf{B}) position ($z$) along the shock propagation direction. We also report the laser pulse shape (laser power is represented in arbitrary units) and vertical lines illustrating the timing and duration of the x-ray flash used to capture one XRD snapshot per experiment. The target package includes a $\sim$215 $\mu m$ ablator, a $\sim$80 $\mu m$ diamond sample and a $\sim$120 $\mu m$ thick  polystyrene reference plate. Solid lines represent the position of the ablator/sample and sample/window interfaces. Here we use the radiative-hydrodynamic simulations \textsc{Hyades} code\cite{larsen1994} with a pressure source scaled to the measured laser power.}\label{fig:Hyades_32645}	
  \end{figure}  
  
  We illustrate the analysis methodology with data from a recent study\cite{millot2025} at Omega EP aiming at documenting the atomic structure of shock compressed diamond near 700 GPa. A $\sim$ 200 $\mu m$ thick Be ablator provides efficient ablation pressure to launch a strong shock into a $\sim$ 100 $\mu m$ thick, 2 mm disk of microcrystalline diamond grown by chemical vapor deposition (Diamond Materials GmbH.) with a 0.7 $\mu m$ Au preheat and photo-ionization shield between the Be and the diamond sample and a $\sim$ 120 $\mu m$ thick polystyrene window. Thanks to the high efficiency of the Beryllium ablator, and to the impedance-mismatch jump the drive laser intensity is near $5\times10^{13}W/cm^2$. Diamond is a low enough Z material and does not absorb enough x-rays from the sample drive ablation plasma plume to induce significant preheat, in particular when shielded by both a thick Be ablator and a Au layer. We note that no significant level of preheat was observed in laser-driven ramp compression experiments of metals\cite{coppari2022}. For an in-depth study of sample preheat by the x-ray source emission see  \cite{krygier2022a}.

  The target package containing the diamond sample is assembled with thin ($\sim$ 1 $\mu m$ ) epoxy bonds and centered on a 400$\mu m$ diameter pinhole on a 10 x 10 mm, 50 $\mu m$ thick platinum foil. This assembly is then positioned and secured on the front face of a PXRDIP box (see Fig.~\ref{fig:Visrad1}). A $\sim$6 $\mu m$ thick Ge film deposited onto 2 x 2 mm, $\sim$ 100 $\mu m$ thick graphite substrate serves as secondary target to produce a x-ray flash for the x-ray diffraction measurements. To filter out a significant fraction of the soft x-ray spectrum generated by the drive ablation plasma we insert 125 $\mu m$ of black Kapton and 100 $\mu m$ of aluminum layers on top of the image plates which line the interior of the PXRDIP box (see Fig.~\ref{fig:PXRDIP_FIlters}).
 
 We use one-dimensional radiative-hydrodynamic simulations with the \textsc{Hyades} code\cite{larsen1994} to design the target assembly and the associated laser pulse shape. Fig.~ \ref{fig:Hyades_32645} provides two-dimensional color maps of the simulated longitudinal stress for the shock experiment on diamond, either in Lagrangian, or in Eulerian coordinates. As a reminder, Lagrangian coordinates follow the fluid individual material elements as they move through space and time, effectively \textit{labeling} each parcel of material, so that the boundary between the various materials appear as straight horizontal lines in the Lagrangian $z-t$ diagram in Fig.~\ref{fig:Hyades_32645}\textbf{A}. In contrast, Eulerian coordinates fix the reference frame in space, describing how fluid properties change at specific spatial locations as material flows through them. The Eulerian $z-t$ diagram in Fig.~\ref{fig:Hyades_32645}\textbf{B} illustrates that phase boundaries begin to move when the compressed material is put in motion behind the shock front. Following the motion of the interfaces (solid dark lines near z=200 and z=300 $\mu m$ at $t=0$) illustrates the approximately two-fold compression of the sample by the shock compression.
  
 A strong shock wave near $450$ $ GPa$ propagates from the Be ablator surface and is transmitted into the diamond sample after $\sim$10 ns, before reaching the polystyrene plate at $\sim$ 14 ns. The 1D hydro simulations suggest that the pressure should be homogeneous to better than ~5\% within the sample during the x-ray flash. 
 In the experiment reported here the VISAR data (not shown) record the arrival and breakout time of the shock wave in the initially transparent diamond sample from which the transit-time velocity can directly be inferred using only the known thickness of the diamond sample plate.  Analysis of the VISAR data suggest that drive non-planarity contributes to less than 5\% variation in pressure.  
 
\begin{figure*}[!t]
 
 	{\includegraphics[width=.89 \textwidth,trim=20 10 70 0 00,clip]{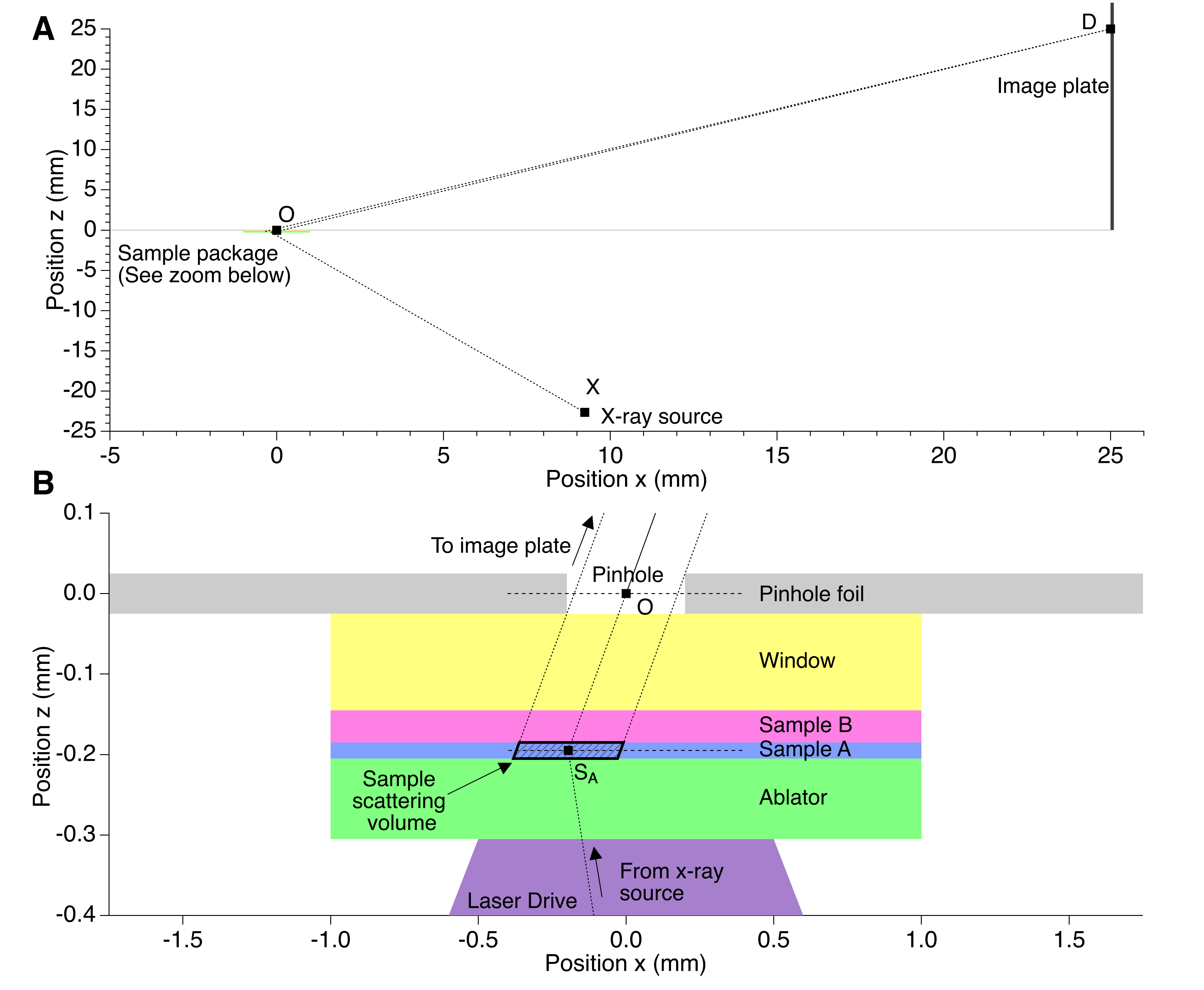}}         
 	
 	\caption{\textbf{PXRDIP x-ray diffraction geometry during the x-ray snapshot.} \textbf{A}: Global view. \textbf{B}: Expanded view around the sample package. These sketches are drawn approximately to scale for the example dataset (note the different magnification along x and z). Determining the position and thickness of the sample layers during the x-ray snapshot requires accounting for the compression and acceleration imparted by the laser drive. For the example discussed in this article, Sample A is the shocked diamond, Sample B in the unshocked diamond ahead of the shock front and the pinhole foil is made of Pt. For a given point on the image plate detector \textbf{D}, only the x-rays diffracted within the {Sample scattering volume} around \textbf{$S_A$} will contribute to the XRD pattern of the compressed sample. }\label{fig:GeoSketchToScale}	
 \end{figure*}

\begin{figure}[!t]
 	\centerline{\includegraphics[width=.49 \textwidth,trim=40 20 110 0,clip]{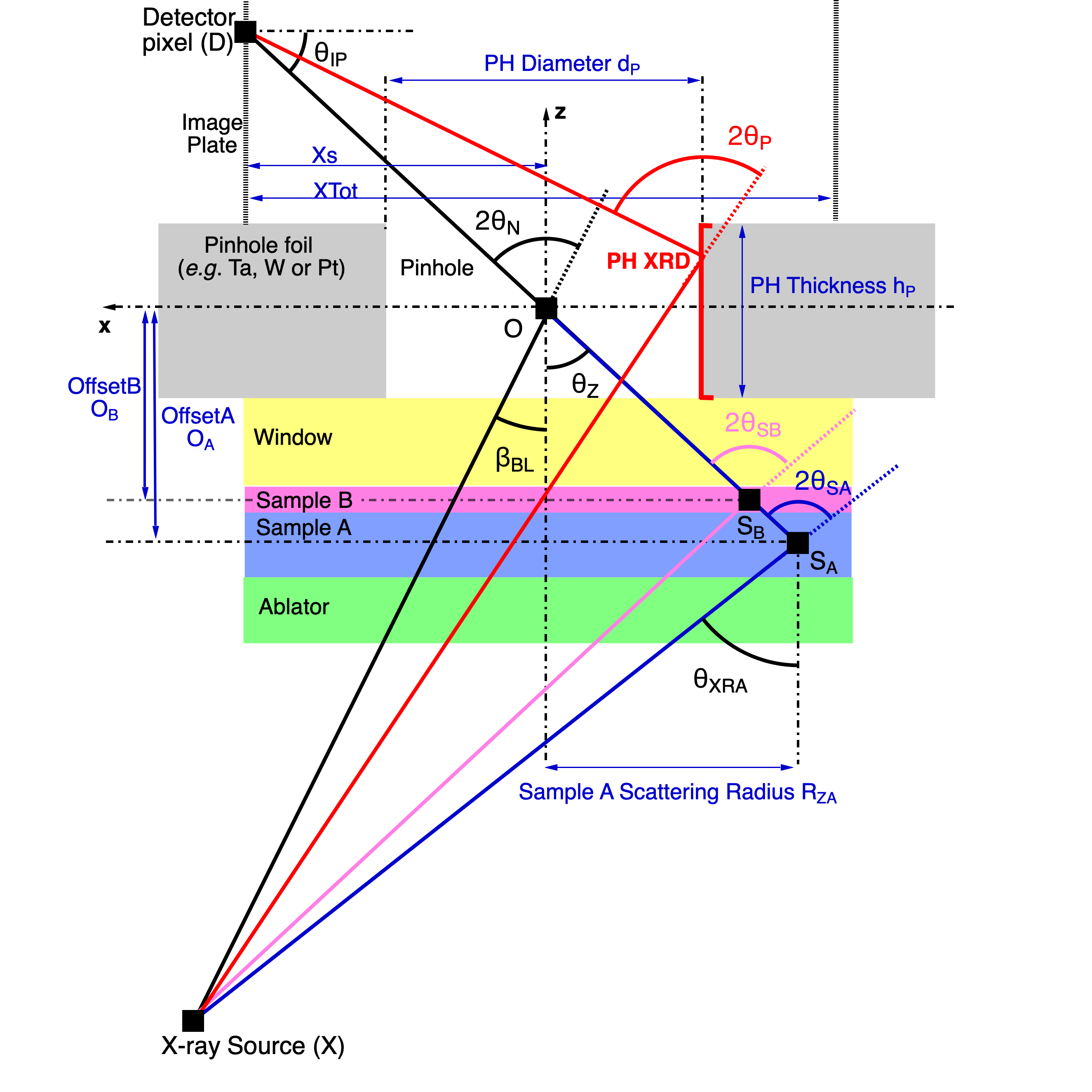}}     
 	\caption{\textbf{Expanded sketch of the PXRDIP x-ray diffraction geometry (not drawn to scale).} We illustrate the various quantities used to model the geometry of the PXRDIP platform for accurate data analysis, including the position of the center of the x-ray source \textbf{X}, the origin of the coordinate system \textbf{O} at the center of the pinhole, and an example point on the image plate detector \textbf{D}. The scattered x-rays from the {Sample scattering volume} (see Fig.~\ref{fig:GeoSketchToScale}) can be accurately approximated \cite{rygg2012,rygg2020} as originating from the sample layer's midpoint $\mathbf{S}_A$ along the \textbf{DO} line of sight. Similarly, contribution from the another Sample B layer can be approximated as originating from $\mathbf{S}_B$.  We use diffracted x-rays from the uncompressed pinhole foil (illustrated in red and collectively labeled PH XRD) as an \textit{in-situ} reference to calibrate the diffraction geometry for each shot.  }\label{fig:GeoSketchNotToScale}	
 \end{figure}

    Here we introduce two parameters used to scale the XRD intensity of the shocked and unshocked fraction to the relevant volume. $d_S$ is the Lagrangian thickness of the shocked region during the XRD snapshot integration time while $d_{U}$ is the thickness of unshocked region. These parameters and the average distance between the shock compressed sample and the center of the platinum pinhole aperture ($O_A$ see Figs.~\ref{fig:GeoSketchToScale} and \ref{fig:GeoSketchNotToScale}) can be determined either directly from the VISAR data in the case of a single shock compression as discussed here, or by interrogating the output of the simulations in Lagrangian coordinates.  In ramp-compression experiments where free-surface velocity is measured with VISAR, $O_A$ can be determined by the characteristics analysis\cite{coppari2022}.
    
    Analysis of the VISAR data for experiment 32645 indicates $d_S=35~\mu m$ and $d_{U}=45~\mu m$ which we use to determine the offset ($O_A$) and ($O_B$) defined in Fig.~\ref{fig:GeoSketchNotToScale} which represent the Eulerian position along the \textbf{z} axis of the mid-point of the shocked and unshocked fractions of the sample during the XRD snapshot.
    
  \section{PXRDIP data analysis workflow}\label{sec:PXRDIP data analysis workflow}

  \begin{figure*}[!htb]
   
  	\centerline{\includegraphics[width=\textwidth,trim=18 265 380 90,clip]{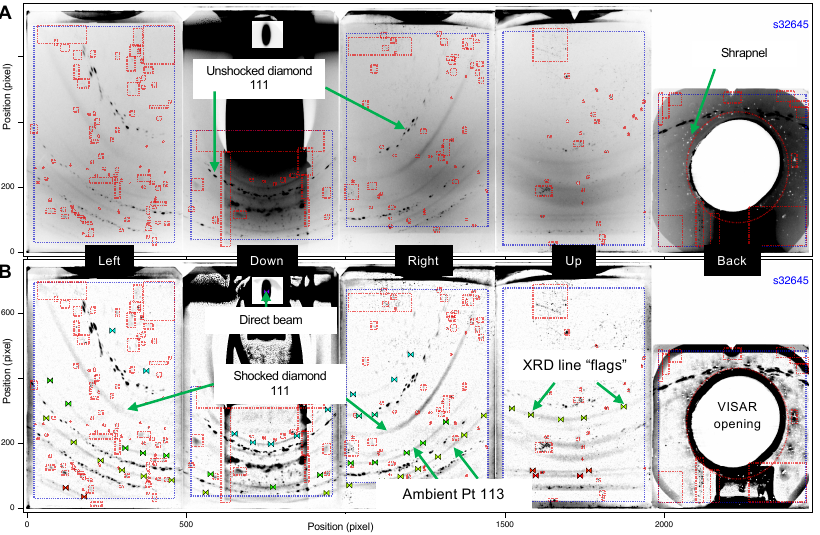}}   
  	 
  	\caption{\textbf{Combined x-ray diffraction image plate scans for 32645 (pixel units, 1 pixel=$100 \mu m$).} (\textbf{A}) As scanned, after cropping and combining. (\textbf{B}) After 2D SNIP background subtraction (using circle kernels, for visualization only, quantitative analysis is performed with vertical line kernels in the $2\theta-\phi$ projection). Data in the vicinity of the direct beam are from the second scan because the first scan is saturated. Color scale is reversed (black is most intense signal). Vertical axis for the Left, Down, Right and Up panels is aligned with the z axis of the PXRDIP box. Back panel is parallel to the sample package. Data outside the blue-dashed boundary for each image plate or inside red-dashed rectangles or ellipses are discarded \textit{i.e.} set to \textit{Not a Number} (NaN) when computing the $2\theta-\phi$ projections. Three spotty diffraction lines from unshocked diamond and the 111 line for shock compressed diamond are clearly visible, together with several uniform lines from ambient platinum. User-defined ``flags" shown as colored bow-ties are used to identify the position of the ambient platinum lines and determine the relative location of the source, sample and detector panels through an iterative calibration procedure.}\label{fig:s30974_Graph_IP}	 
  \end{figure*}  
  
  \subsection{Geometry and raw data}\label{Sec:Geometry and data format}
  After each experiment, the PXRDIP box is recovered from the target chamber and the Fujifilm BAS-MS image plates (IPs) are retrieved and scanned  with a General Electric Typhoon 7000FLA scanner approximately $30 \pm10$ min after the experiment, with a sensitivity of 1000 and a 100 $\mu m$ pixel size. 
  
Figs.~\ref{fig:Visrad1}, \ref{fig:GeoSketchToScale} and \ref{fig:GeoSketchNotToScale} illustrate the geometry for Powder X-Ray Diffraction Image Plate (PXRDIP) experiments at the Omega EP Laser Facility. X-rays are emitted in all directions from the x-ray source\cite{coppari2019} and illuminate the entire sample package. Placing a $300-500~\mu m$ diameter pinhole aperture behind the sample stack and the IP detector about 30 mm away strongly limits the angular distribution of the direct and diffracted x-rays that can reach the image plate detectors. As shown in Fig.~\ref{fig:GeoSketchToScale}, only the x-rays diffracted within the {Sample scattering volume} around the point $\mathbf{S}_A$ contribute to the XRD pattern at a given point on the image plate detector \textbf{D}. This contribution can be approximated \cite{rygg2012,rygg2020} as originating from the sample layer's midpoint \textbf{$\rm S_A$} (Fig.~\ref{fig:GeoSketchNotToScale}) along the \textbf{DO} line of sight (Fig.~\ref{fig:GeoSketchToScale} \textbf{A}), \textit{i.e.} offset towards \textbf{-z} by $O_A$ from the center of the pinhole \textbf{O}. Ray-tracing calculations (not shown) confirm that this approximation is still valid for a shocked sample having a  compressed thickness near $50-100~\mu m$ during the XRD snapshot. Similarly, contribution from the unshocked diamond layer can be approximated as originating from $\mathbf{S}_B$ in Fig.~\ref{fig:GeoSketchToScale}. 

 For each point \textbf{D} on one of the image plate panels, we aim at determining  the diffraction angle ($2\theta$) and the azimuth angle around the direct transmitted beam ($\phi$) (we set $\phi=0$ for diffracted x-rays towards the z axis) to project the data into $2\theta-\phi$ (see Section~\ref{sec:Projection into 2TT-phi}). Computing the \textit{Scattering Radius} ($R_{ZA}$) \textit{i.e.} the radius of the center of the {Sample scattering volume} is also useful to evaluate the impact of drive non-uniformity (see Section~\ref{sec:Projection into 2TT-radius}). 
 As illustrated in Fig.~\ref{fig:GeoSketchNotToScale} $R_{ZA}=O_A~\rm{ sign}(\phi)~tan~\theta_Z$, where $O_A$  is the offset along the z axis,  $\theta_Z$ is the attack angle \textit{i.e.} the angle between sample normal and the diffracted x-ray and $\rm sign(\phi)=1$ if $\phi\geq0$ and -1 otherwise. It is also useful to compute the incidence angle of the diffracted rays onto the image plates ($\theta_{IP}$) to determine the IP efficiency and the attenuation of the filters inserted inside the box. Similarly, we calculate the angle between the normal to the sample and the incident ray to $S_A$ ($\theta_{XRA}$) to determine the attenuation induced by the ablator. 
  
We use diffracted x-rays from the uncompressed platinum foil as an \textit{in-situ} reference to determine the relative position of the source, sample package and of the various image plate panels for each shot. For each point \textbf{D}, we model this signal as originating from portions of the walls and rims of the pinhole (illustrated in red and collectively labeled PH XRD in Fig.~\ref{fig:GeoSketchNotToScale}) which have a direct view -- or a quasi-direct view \textit{i.e.} through less than one attenuation length -- of both \textbf{X} and \textbf{D} using the approach described in Ref.\cite{rygg2020} and the corrected expression\cite{rygg2026}.

In the experiment described here we only use the pinhole lines to perform the geometry calibration.

 As shown on Fig.~\ref{fig:s30974_Graph_IP}, we collect XRD data onto five image plate panels lining the walls of the PXRDIP box (Fig.~\ref{fig:PXRDIP_FIlters}b). All panels are scanned together once to produce the main dataset. A second scan is then collected to enable distinguishing the position of the direct beam without saturation.

  \begin{figure}[!htb] 
 
  	\centerline{\includegraphics[width=.43\textwidth,trim=30 30 30 70,clip]{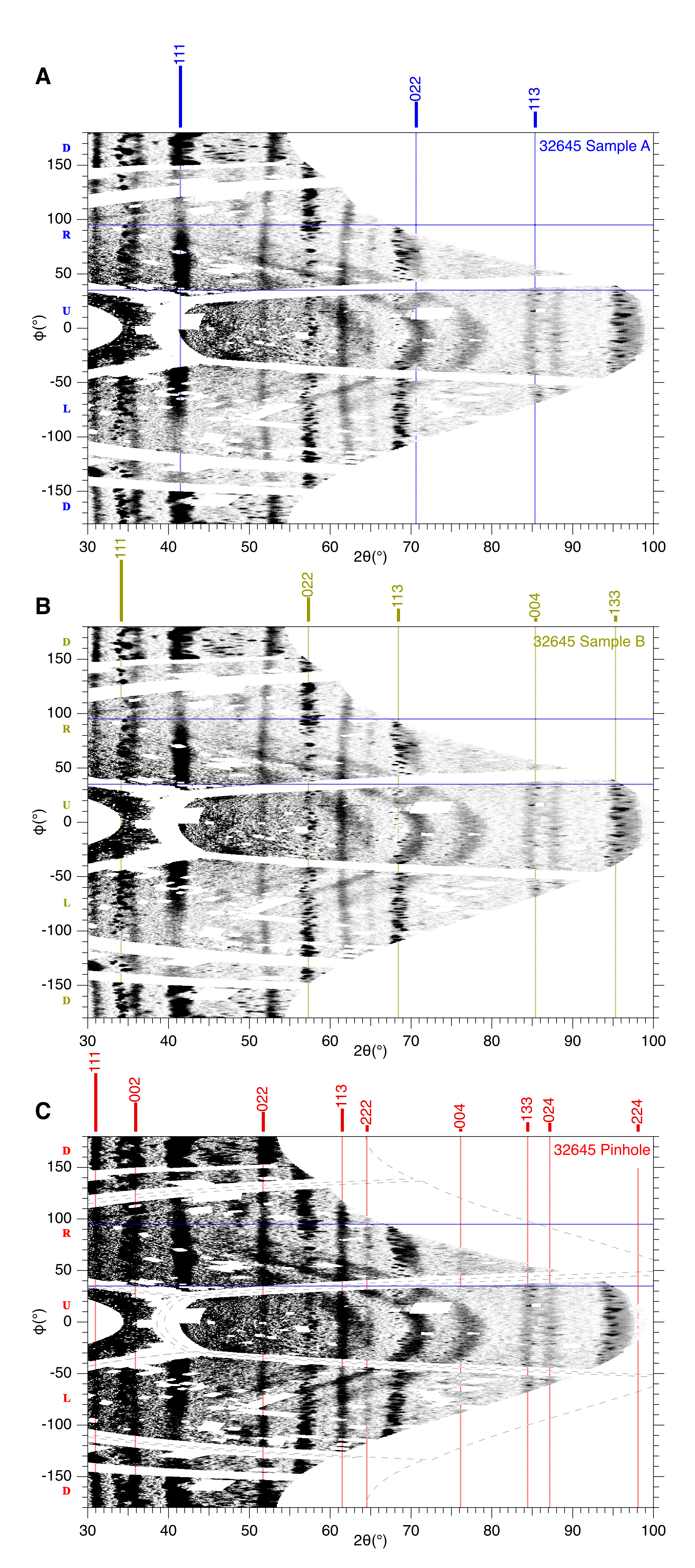}}   
  	\caption{\textbf{Background subtracted and cropped XRD data for 32645 projected into azimuth angle versus scattering angle ($\phi-2\theta$).} Vertical ticks and Miller indices represent the theoretical XRD pattern of compressed diamond ($a=2.959\rm\AA$, blue), ambient diamond ($a=3.567\rm\AA$, dark yellow) and ambient platinum ($a=3.923\rm\AA$, red). (\textbf{A}) Sample A (shocked diamond) projection: we observe the 111 line. (\textbf{B}) Sample B (ambient diamond) projection: we observe the 111, 022, 113 and 133 lines. (\textbf{C}) Pinhole (ambient Pt) projection: we observe the 111, 002, 022, 113, 222, 133 and 024 lines.  Color scale is linear, reversed (black is most intense signal). Grey dashed lines in (\textbf{C}): projected contours of IP panels L, D, R, U and B. Blue horizontal lines correspond to the thick blue lines on Fig.~\ref{fig:SNIPRadius}. The apparent curvature of the Sample B lines in (\textbf{C}) and of the Pt lines on (\textbf{A}) illustrate the difference in scattering location. In addition, different intensity corrections are applied to the three projections to account for the difference in scattering location. As a consequence, the diffraction signal of each layer, should be extracted from the corresponding projection (driven sample intensity from projection \textbf{A}, ambient sample from \textbf{B} and pinhole intensity from \textbf{C}, see Sec \ref{sec:Projection into 2TT-phi}  for details).  }\label{fig:XRD_MCD_OutPut_32645}	 
  \end{figure} 
 
    \begin{figure*}[!htb]
  	 
  	\centerline{\includegraphics[width=.89
  		\textwidth,trim=0 0 0 0,clip]{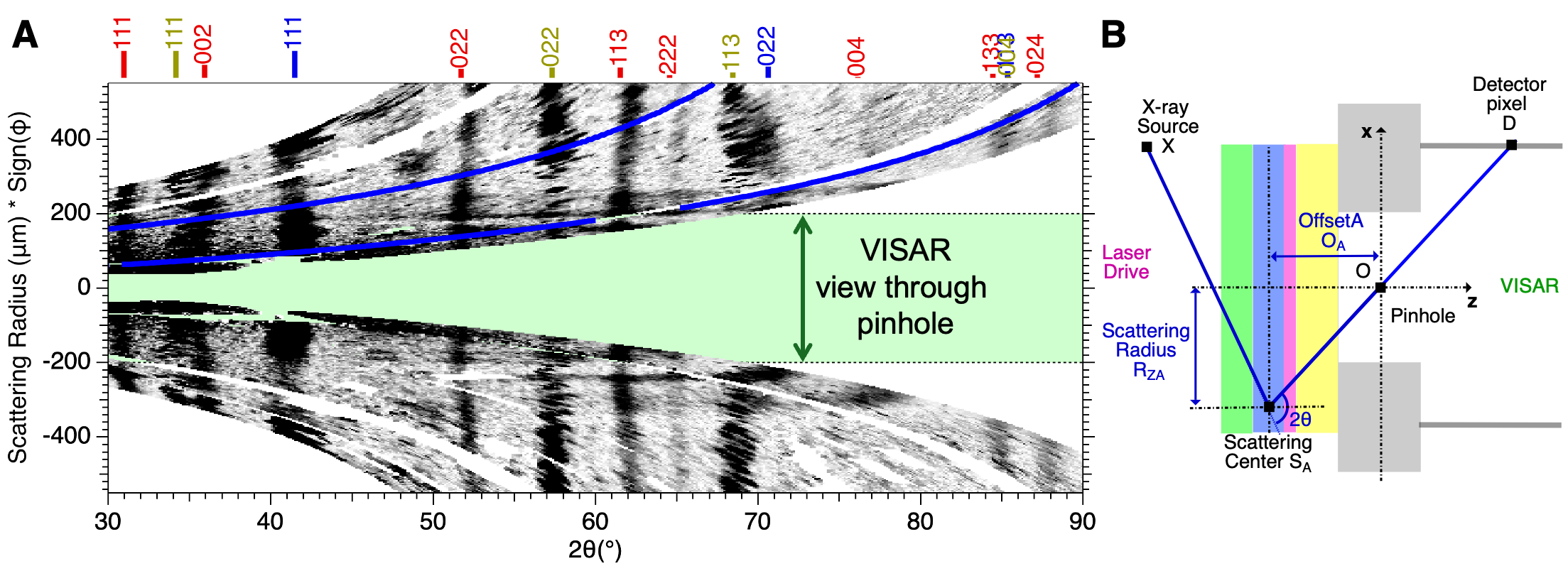}}   
  	\caption{\textbf{Scattering Radius projection.} \textbf{A:} X-ray diffraction data for 32645 projected into Scattering Radius ($R_{ZA}$) versus diffraction angle ($2\theta$). The thick blue lines delimit the range of radii corresponding to the range $35^{\circ}< \phi<95^{\circ}$ between the blue horizontal lines in Fig.~\ref{fig:XRD_MCD_OutPut_32645}\textbf{A} and illustrate that data at constant $\phi$ do not originate from the same radius in the sample mid-plane. Vertical ticks with Miller index labels represent the theoretical XRD pattern of ambient platinum (red), ambient diamond (dark yellow) and compressed diamond (blue). The green shaded area represents the extent of the pinhole, \textit{i.e.} of the diameter of the area over which the spatial and temporal uniformity of the compression can be monitored with the line-imaging VISAR. \textbf{B:} Simplified geometry sketch (see Fig.~\ref{fig:GeoSketchNotToScale}) illustrating the definition of the \textit{Scattering Radius} ($R_{ZA}$) as the distance of the center of the Sample A \textit{Scattering Volume} ($S_A$) from the \textbf{z} axis.}\label{fig:SNIPRadius}	 
  \end{figure*}  
  
  \subsection{Projection into $2\theta-\phi$}\label{sec:Projection into 2TT-phi}  
  \textsc{AnalyzePXRDIP} loads the images in the native .gel format and converts the scanned intensity of each pixel $G$ to obtain the intensity in linearized photostimulated luminescence units\cite{williams2014} ($IP_{PSL}$) with:
  \begin{equation}
  	IP_{PSL}=\left(\frac{G}{(2^B-1)}\right)^2~\left(\frac{R}{100}\right)^2
  	~\left(\frac{4000}{S}\right)~10^{(L/2)} 
  \end{equation}
  where the spatial resolution in $\mu m$ $R=100$, the sensitivity $S=1000$, the latitude $L=5$ and the bit depth $B=16$. We then crop and rearrange the panels into a combined IP image for each of the first and second scans. We then overwrite the saturated data around the direct beam on the combined IP first scan image with data from the second scan and obtain a composite, combined IP image used to compute the projections into $2\theta - \phi$ and perform quantitative analysis of the XRD data. This is illustrated in Fig.~\ref{fig:s30974_Graph_IP}\textbf{A} where the portion of the image from the second scan is clearly visible.
  
Assuming the positions of the x-ray source, sample package and individual image plate panels, and using the the x-ray source wavelength, we can analytically compute the $2\theta$ and $\phi$ angles corresponding to each detector point and project the image plate data onto $2\theta_{Pin}-\phi_{Pin}$, $2\theta_{SA}-\phi_{SA}$ and $2\theta_{SB}-\phi_{SB}$ as illustrated on Fig.~\ref{fig:XRD_MCD_OutPut_32645}. Here we use subscript labels to identify the  diffraction and azimuth angles for the Pinhole signal and the contributions from Sample A originating from $S_A$ and that from Sample B originating from $S_B$. In the following method description we drop these labels for clarity.  

Projecting the data onto $2\theta$ and $\phi$  is not done by a global analytical warping function because of the unusual geometry of the PXRDIP source-sample-collimating aperture-detector arrangement. Instead, we define a new, 1000$\times$1000 2D array $P(2\theta-\phi)$ to represent the projected data into $2\theta-\phi$. Columns represent $\phi$ while lines represent $2\theta$. We determine the intensity of the projected image $P(2\theta-\phi)$ by first determining the $2\theta$ and $\phi$ value which correspond to the center of each pixel on the image plates, then accounting for the fact that multiple IP pixels may project onto the same $2\theta-\phi$, and finally using a multi-pass median filtering algorithm to fill in the $2\theta-\phi$
  pixels that are not mapped to one or more IP pixels.

  Due to the single-shot nature of our experimental approach -- both the sample and the secondary target are destroyed after the laser pulse, and their positions with respect to the detector and laser beams vary for each experiment -- we cannot rely on the use of a known x-ray diffraction reference material  such as $\rm CeO_2$ powder positioned at the nominal position of the sample to determine the XRD geometry. Instead we use an iterative calibration approach, identifying features on the image plate data and a minimization algorithm to determine the respective position of the x-ray source, of the sample package and of each of the image plate panels. 
  
  We begin by assuming that the x-ray source and the sample are at their nominal position in the target chamber, that the PXRDIP box was machined perfectly and that the image plates are at their nominal position relative to the walls of the box. We then compute the projections into $2\theta-\phi$ before applying a 2D Statistics-sensitive Non-linear Iterative Peak-clipping (SNIP) algorithm to estimate and subtract the background\cite{rygg2020,morhac2008,morhac2009}. Here we use a 2D SNIP with constant $2\theta$ lines and manually position a number of \textit{flags} to mark the position of known features either on the  $2\theta_{Pin}-\phi_{Pin}$ projection or directly on the image plate panels.
  In the example discussed here we used the position of the direct beam as well as six diffraction lines of the uncompressed platinum (see Figs.~\ref{fig:s30974_Graph_IP}). 
  
 To accurately model the diffraction signal from the pinhole, we use the approach described in Eq.~6 in Ref.\cite{rygg2020}. For each detector point \textbf{D} we compute a diffraction angle ($2\theta_P$) for the signal originating from the pinhole walls and rims. To do this we divide the pinhole walls and rims into small volume elements and compute the diffraction angle ($2\theta_i$) for each element having a direct view of both the source \textbf{X} and the detector \textbf{D}. A weighted average\cite{rygg2020} of all $2\theta_i$ at each point \textbf{D} then yields $2\theta_P$. 
  
  Similarly to early work\cite{kalantar2005}, we then use a minimization algorithm to find the optimum set of distances and angles describing the source and sample positions and the position and rotation of each IP panel. This new set of parameters is then used to compute the projections into $2\theta-\phi$ and this process is iterated until convergence, typically when the discrepancy between the expected and observed $2\theta$ values are less than 0.3 degrees and no significant curvature is apparent for the various diffraction lines of the uncompressed pinhole on the $2\theta_{Pin}-\phi_{Pin}$ projection. Because the 2D SNIP background subtraction uses constant $2\theta$ lines as kernel, its accuracy and the ability to retrieve weak signals despite the intense background will improve as the iterative process to determine the diffraction geometry approaches convergence and the pinhole, Sample A and Sample B diffraction lines appear "vertical" in their respective $2\theta-\phi$ projection. 
    
    It is often useful to mask spurious features including shrapnel and Laue diffraction spots before projecting into $2\theta-\phi$ and performing the SNIP background subtraction. This is illustrated with the red dashed rectangles in Fig.~\ref{fig:s30974_Graph_IP}.   
    
  \subsection{Projection into $2\theta-\rm R_{ZA}$ and connection with line-imaging VISAR}\label{sec:Projection into 2TT-radius}

 Projecting the IP data as a function of the {Scattering Radius} ($R_{ZA}$) versus diffraction angle ($2\theta$) is a convenient way to identify fingerprints of potential deviation from a 1D uniform compression. Following the projection of constant $\phi$ lines (blue curves in Fig.~\ref{fig:SNIPRadius}) into $2\theta-\rm R_{ZA}$ illustrates that PXRDIP and TARDIS data at constant $\phi$ do not originate from $S_A$ points located at the same $R_{ZA}$. In other words, the integrated PXRD pattern as a function of $2\theta$ is a composite pattern. This is a characteristic of the PXRDIP/TARDIS geometry with a collimating pinhole positioned behind the sample. This particularity is important to keep in mind if the sample is not in a uniform strain state or if preferred orientation is present. For example, small drive non-planarity may induce non-uniform strain in the sample resulting in tilted or curved lines in $R_{ZA}-2\theta$.  

At Omega and EP, the VISAR diagnostic used to document the compression history is a line-imaging system, \textit{i.e.} it records the velocity history for a series of point along its {imaging direction}. The {imaging direction} corresponds to the projection of the streak camera entrance slit onto the target plane\cite{celliers2023}. Comparing the orientation of the VISAR {imaging direction} with the positioning of the PXRDIP box (see  Fig.~\Ref{fig:Visrad1}) reveals that the standard configuration of the VISAR system aligns the {imaging direction} with the x-axis. This means that the VISAR documents potential differences (\textit{i.e.} drive non-planarity) between the Up and Down PXRDIP panels (see Fig.~\ref{fig:s30974_Graph_IP}). Because the Left and Right panels cover a nearly symmetric, large range of $2\theta-\phi$ space, whereas the Up panel only covers a small $\phi$ range, it may be beneficial to rotate the {imaging direction} by 90 degrees to document drive asymmetries between the Left and Right panels. This could be accomplished by rotating a Dove prism located between the interferometer and the streak camera.

  \section{XRD Intensity corrections before background subtraction}\label{sec:XRD Intensity corrections before background subtraction}
 
    In addition to determining the diffraction angle ($2\theta$) of XRD lines for the Sample A and Sample B projections, we aim at determining the relative intensity of the diffraction signal from the Sample A and B and from the pinhole (see Fig.~\ref{fig:s30974_Graph_IP}). Because these various signals are attenuated differently before reaching the detector, our data analysis workflow implements several steps specific to the PXRDIP geometry. We apply specific, pixel-dependent intensity corrections to the three $2\theta-\phi$ projections. We implement both new and previously described corrections\cite{rygg2012,rygg2020} using notations from Figs.~\ref{fig:Visrad1}, \ref{fig:GeoSketchToScale} and \ref{fig:GeoSketchNotToScale}. Here we continue discussing the example dataset on shock compressed diamond for which Sample A is the shocked diamond, Sample B is the unshocked diamond and the pinhole reference is Pt.

 \subsection{Effective detector pixel area} \label{sec:Effective detector pixel area}
 In contrast with a standard powder XRD spectrometer, the sample to detector distance is not uniform across the various IP panels. We account for the effective area of the IP pixel by correcting the intensity of the image ($IP(p,q)$) at pixels \textit{p, q} following Eq. 45 in Ref.\cite{rygg2020}:
  \begin{equation}
  	Int_1(p,q)=IP(p,q) ~( \rm r_D/r_{D0})^2/cos ~\theta_{IP} 
  \end{equation}
  where $r_D$ is the distance of the pixel \textit{(p,q)} of the IP panel from the origin of the coordinate system \textit{i.e.} the norm of \textbf{OD}, $r_{D0}$ is a scaling factor on the order of the average sample-detector distance over the full set of IPs (for PXRDIP we use $r_{D0}=500$ pixels or 50 mm) and $\theta_{IP}$ is the incidence angle of the diffracted rays onto the IP detector (see Fig.~\ref{fig:GeoSketchNotToScale}). This correction is applied to the Sample A, Sample B and Pinhole projections.
  
   \subsection{Pinhole sample-area detected}\label{sec:Pinhole sample-area detected}
   In the PXRDIP/TARDIS geometry the size of the sample scattering volume is not uniform for all points \textbf{D} on the image plates (see Fig.~\ref{fig:GeoSketchToScale}). The observed intensity will therefore vary significantly, proportionally to the sample scattering volume for each detector point \textbf{D}. Given the diameter ($d_p$) and the height ($h_p$) of the pinhole (see Fig.~\ref{fig:GeoSketchNotToScale}) and the attack angle ($\theta_Z$), Eq. 43 in Ref.\cite{rygg2020} proposed using the pinhole effective area $A_{Eff}$ as a scaling factor. However (see also Ref.\cite{rygg2026}), the relevant correction quantity is the sample-area detected ($A_{Det}$):
   \begin{align}
   	A&_{Det}(p,q)=A_{Eff}(p,q)~sec~\theta_Z
   \end{align} 
   Using the attack angle $\theta_Z(p,q)$ at each pixel on the image plate \textit{p, q} and defining $J=tan~\theta_Z~ h_p/d_p$  we obtain $A_{Det}$ for each detector point D :
   \begin{align}
  A&_{Det}=\frac{d_p^2}{2}~\left[\frac{\pi}{2}-\arcsin(J)-J~cos~(\arcsin(J))\right]	  
    \end{align} 
    
 The portions of the detectors with very small $A_{Det}$, \textit{i.e.} near the boundary of the $2\theta-\phi$ coverage will therefore have very attenuated XRD signal from the sample. This often results in low signal-to-noise and signal-to-background data. 
 To avoid including these areas in the intensity lineouts and optimize the contrast to better visualize the data, we usually set the regions of the IPs where the sample-area detected $A_{Det}$ is smaller than a cutoff value (user-defined for each experiment, for example to 10-20\% of the area of the pinhole aperture) to NaN and correct the intensity elsewhere by the sample-area detected factor normalized to the area of the pinhole under normal incidence \textit{i.e.} $A_{Det}/(\pi~d_p^2/4)$. We correct the Sample A  and Sample B projections with:
  \begin{equation}
  	Int_2(p,q)=\frac{Int_1(p,q)}{A_{Det}/(\pi~d_p^2/4)}
  \end{equation}	
  
    \subsection{Intensity correction for the lines of the pinhole}\label{sec:Intensity correction for the lines of the pinhole}
 Modeling the XRD signal from the pinhole is not trivial because the view factor of the source and detector vary for different points of the walls and rims. In section \ref{sec:Projection into 2TT-phi} we summarize the approach described in Ref. \cite{rygg2020} to compute the average diffraction angle ($2\theta_P$). Here we propose that the total volume of pinhole material contributing to the diffraction at each point \textbf{D} on the image plate detectors can serve as an estimate of the expected pinhole XRD signal intensity. Users can therefore optionally correct the intensity of the pinhole projection as:
  \begin{equation}
  	Int_{2P}(p,q)=Int_1(p,q)~C/V_{0}
  \end{equation}	
  where $C$ is arbitrary scaling factor and $V_0$ is the volume integral factor defined in Eq. 16 in Ref. \cite{rygg2020}. $C$ is user-defined for each experiment. Here we use $C =2\times 10^4$ to ensure that this correction is about 1 near the center of the image.  
  
  Here we assume that the  crystallographic texture in the pinhole foil does not affect the measured diffraction intensity. We note that we expect a good powder statistics because the pinhole signal at each detector point is the sum of diffraction from various locations around the $300-500\mu m$ diameter pinhole, \textit{i.e.} over a large number of crystallites. 
  
  \section{SNIP background subtraction and further intensity corrections}\label{sec:SNIP background subtraction and further intensity corrections}
We use the 2D SNIP algorithm\cite{rygg2020,morhac2008,morhac2009}  with constant $2\theta$ kernels to perform a background subtraction on each of the three  projections into $2\theta-\phi$ (see Section \ref{sec:Projection into 2TT-phi} and Fig.~\ref{fig:XRD_MCD_OutPut_32645}). Because the angular instrumental resolution depends strongly on the source distance and on the pinhole diameter, it is important to determine what SNIP kernel width is appropriate for each experiment. This can be done by comparing intensity lineouts of the sample $2\theta-\phi$ projection after SNIP subtraction with a range of kernels widths. Too small of a kernel width may clip the XRD peaks too aggressively and reduce their width, thus strongly affecting the determination of the XRD intensity, while large background may subsist when too large of a kernel size is used. Narrow kernels may nonetheless prove useful to retrieve peaks with small signal-to-background ratio.
  
\begin{figure}[!b]
  	\centerline{\includegraphics[width=.49 \textwidth,trim=0 0 0  00,clip]{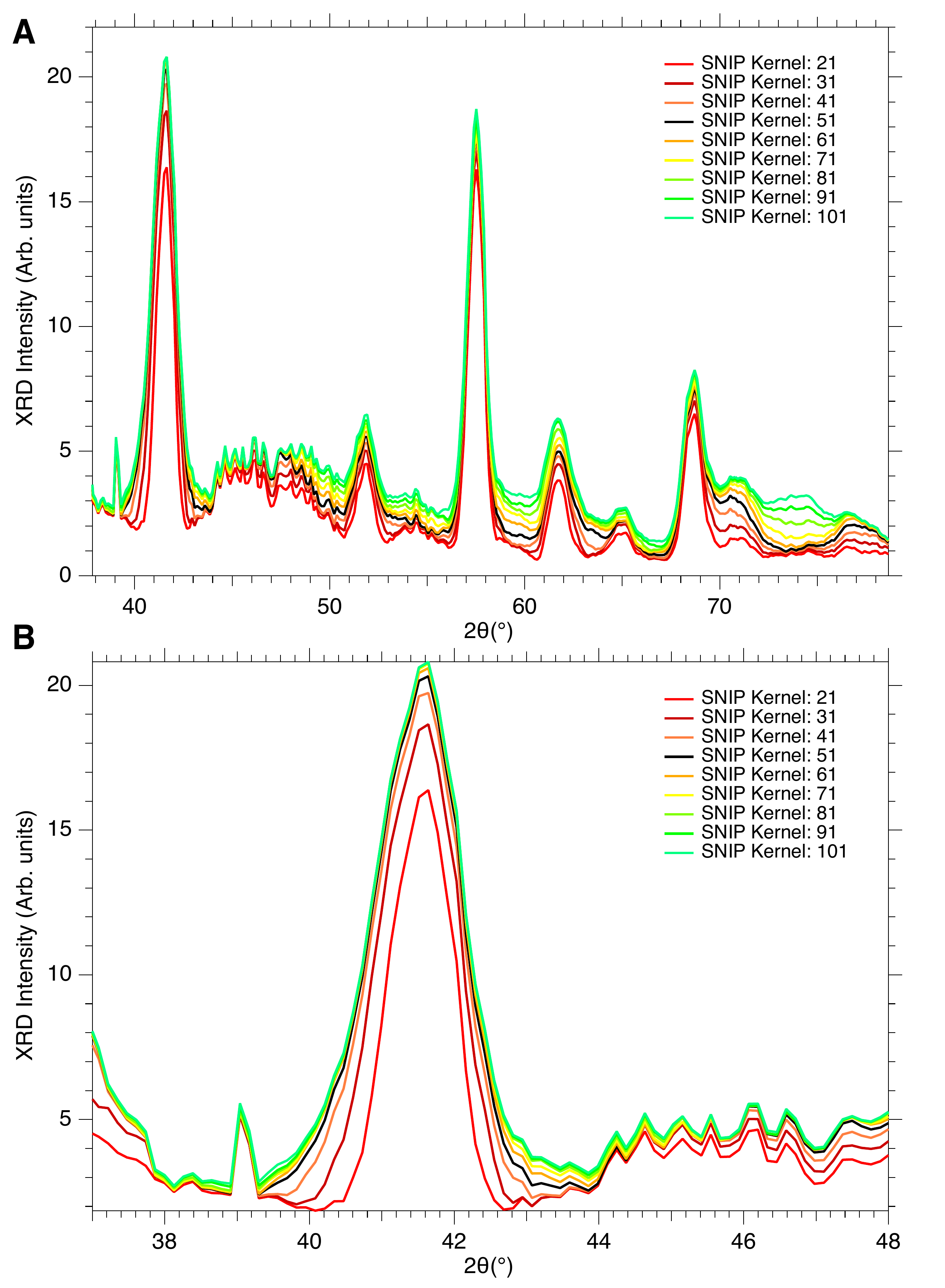}}     
  	\caption{\textbf{Influence of the SNIP kernel width.} Series of lineouts of a portion of the sample $2\theta-\phi$ projection for experiment 32645 with a $400 \mu m$ diameter pinhole: (\textbf{A:}) Broad $2\theta$ range. (\textbf{B:}) Sample peak. The 51 pixel wide kernel (black curve) appears optimum to retrieve the sample peak area to determine the XRD intensity. }\label{fig:CompareSNIP}	
 
  \end{figure}

  In many experiments with a 300 $\mu m$ pinhole and a sample-x-ray source distance near 24 mm, we find that a kernel of 41 pixels is adequate. Fig.~\ref{fig:CompareSNIP} shows a series of lineouts for the sample projection corresponding to SNIP background subtractions with various kernel sizes for the example dataset on shocked diamond. We find that in this case with a 400 $\mu m$ pinhole diameter the 51 pixel wide kernel (black curve) appears optimum to retrieve the sample peak area to determine the XRD intensity.

  Because most of the broadband background generated by the drive and/or x-ray source ablation plasma plumes is removed by the SNIP algorithm, we then apply a few additional intensity corrections for which we assume that the observed signal was generated by diffracted x-rays at the peak energy of the x-ray source\cite{coppari2019}, \textit{e.g} $E=10.249\rm ~keV$ (Ge) or $E=8.3680$ keV (Cu). 
  
\subsection{IP Sensitivity correction}\label{sec:IP Sensitivity correction}
Image plate detectors are very versatile and resilient with a large dynamic range. However, determining their sensitivity is not trivial, in particular when considering the non-normal incidence configuration of the PXRDIP/TARDIS geometry. We follow the approach proposed in Ref.\cite{rygg2020} with two modifications.  First, we use the chemical formula for the phosphor $Ba_{2263}F_{2263}Br_{1923}I_{339}C_{741}H_{1730}N_{247}O_{494}$ and the density $\rho=3.3 g/cm^3$ from Ref.\cite{stoeckl2018} to compute the energy-dependent linear absorption coefficient $\mu_d$ in the phosphor layer of the BAS-MS image plates using known mass attenuation coefficients. We obtain $\mu_d=1/(25.4\mu m)$ at  $E=10.249\rm ~keV$.
Second, we compute the scaled sensitivity of the image plates as a function of incident angle ($\theta_{IP}$) and photon energy ($E$) ($Q(E,\theta_{IP}$)) as :
  \begin{align}
  	Q(E,\theta_{IP})&=\frac{E}{Q_0}~\frac{sec~\theta_{IP}}{(sec~\theta_{IP}+1/(\mu_d~L))}\nonumber\\
  	& \times(1-e^{-\mu_d~h_d~(sec~\theta_{IP}+1/(\mu_d~L))})  
  \end{align}
 where $E$ is in keV, $h_d=115$ $\mu m$ is the thickness of the phosphor layer, the readout length $L=125~\mu m$, $\theta_{IP}$ is the incidence angle of the diffracted rays onto the image plate (see Fig.~\ref{fig:GeoSketchNotToScale}) and $Q_0$ is a scaling factor defined as the IP sensitivity at normal incidence for a Ge x-ray source. This is equivalent to the corrected expression\cite{rygg2026} for Eq.~49 in Ref.\cite{rygg2020}.

 Note that the x-ray photon energy ($E$) enters this formula: the IP sensitivity depends on the energy deposited\cite{rygg2020} and not on the number of photons so that, for a given incident energy flux the photostimulated luminescence signal readout by the IP scanner will be stronger when using a Ge x-ray source, compared to a Cu x-ray source.
 
 To account for the variation of the sensitivity across our IP detector panels we  divide the intensity at each pixel in $2\theta -\phi$ by $Q$, using the peak energy of the x-ray source for $E$. 
 
\subsection{Attenuation from filters, sample package and IP cover layer} \label{sec:Attenuation from filters, sample package and IP cover layer}
We additionally correct the intensity of the signal to compensate for the attenuation due to the filters inside the PXRDIP box (see Fig.~\ref{fig:PXRDIP_FIlters}), the sample package materials and the thin mylar cover layer on the image plates using energy-dependent mass-attenuation factors\cite{henke1993}. 
For each material layer in the sample package or in the PXRDIP box we follow Ref.~\cite{rygg2020} and account for the incidence angle ($\eta$) so that the transmission of a given layer ($T$) is given by $T=T_0^{sec~\eta}$ where $T_0$ is the transmission at normal incidence computed with the thickness, density and mass attenuation of the material constituting the layer.
Because the scattering location for the diffraction signal of the shocked sample, the unshocked sample and the pinhole is slightly different (see Fig.~\ref{fig:GeoSketchNotToScale}), the incidence angle of the incident and diffracted rays and therefore the  x-ray attenuation is also slightly different.
For the attenuation through the Mylar layer we use the composition\cite{stoeckl2018} $C_{10}H_8O_4$, a density $\rho=1.4 g/cm^3$ and a thickness of 9 $\mu m$. For the metallic filters, we assume that the aluminum and copper filters are perfect cylinders oriented along the \textbf{z} axis, so that the incidence is $\pi/2-\theta_Z$. For Kapton we use the composition $C_{22}H_{10}N_2O_5$ and a density $\rho=1.42 g/cm^3$.
We correct the shocked and unshocked sample projections for the absorption in the ablator, shield, shocked sample, unshocked sample and window layers. For the samples A and B we use the incidence angles $\theta_{XRA}$ and $\theta_{XRB}$ (Fig.~\ref{fig:GeoSketchNotToScale}) and $\theta_Z$ as the exit angle to precisely determine the attenuation using Equation 42 in Ref.\cite{rygg2020}. For the pinhole projection we use the incidence angle for the x-ray source at the coordinates origin \textbf{O} at the center of the pinhole. 
  
 \section{Post-processing and  x-ray diffraction intensity analysis}\label{sec:Post-processing and  x-ray diffraction intensity analysis}

  \subsection{Intensity lineouts and fits}\label{Sec:Intensity lineouts and multi-peak fits} 
To enable quantitative analysis of the powder diffraction patterns, it is common practice to compute a lineout of the XRD intensity as a function of $2\theta$. However, the specific PXRDIP geometry renders this step challenging because background features may persist after the SNIP background subtraction and signal-to-noise and signal-to-background ratios may be low. In addition, Sample A, Sample B and pinhole diffraction signals originate from different locations and different corrections must be applied to each of them. As shown on Fig.~\ref{fig:XRD_MCD_OutPut_32645}\textbf{A} pinhole lines appear curved in the Sample A projection while Sample B lines appear curved in the pinhole projection. It is therefore important to compute lineouts to analyze the  Sample A, Sample B and pinhole diffraction patterns in their respective $2\theta-\phi$ projection when comparing the respective intensity of these different contributions.
  
  For the example dataset shown on Fig.~\ref{fig:s30974_Graph_IP}, we obtain the background subtracted projections shown on Figs.~\ref{fig:XRD_MCD_OutPut_32645} and \ref{fig:SNIPRadius}. We compute three XRD intensity lineouts by averaging the background subtracted projection over a common range of $\phi$ for each of the three $2\theta-\phi$ projections, after masking spurious features (see Fig.~\ref{fig:XRD_MCD_OutPut_32645}). While the lineout would ideally be averaging the full extent of the $\phi$ coverage enabled by our diffraction geometry, the presence of strong background due to the hot ablation plasma may significantly affect the signal-over-background ratio in some portions of the image plates. 
  
\begin{figure}[!ht ]
  	\centerline{\includegraphics[width=.49\textwidth,trim=0 0 0 0,clip]{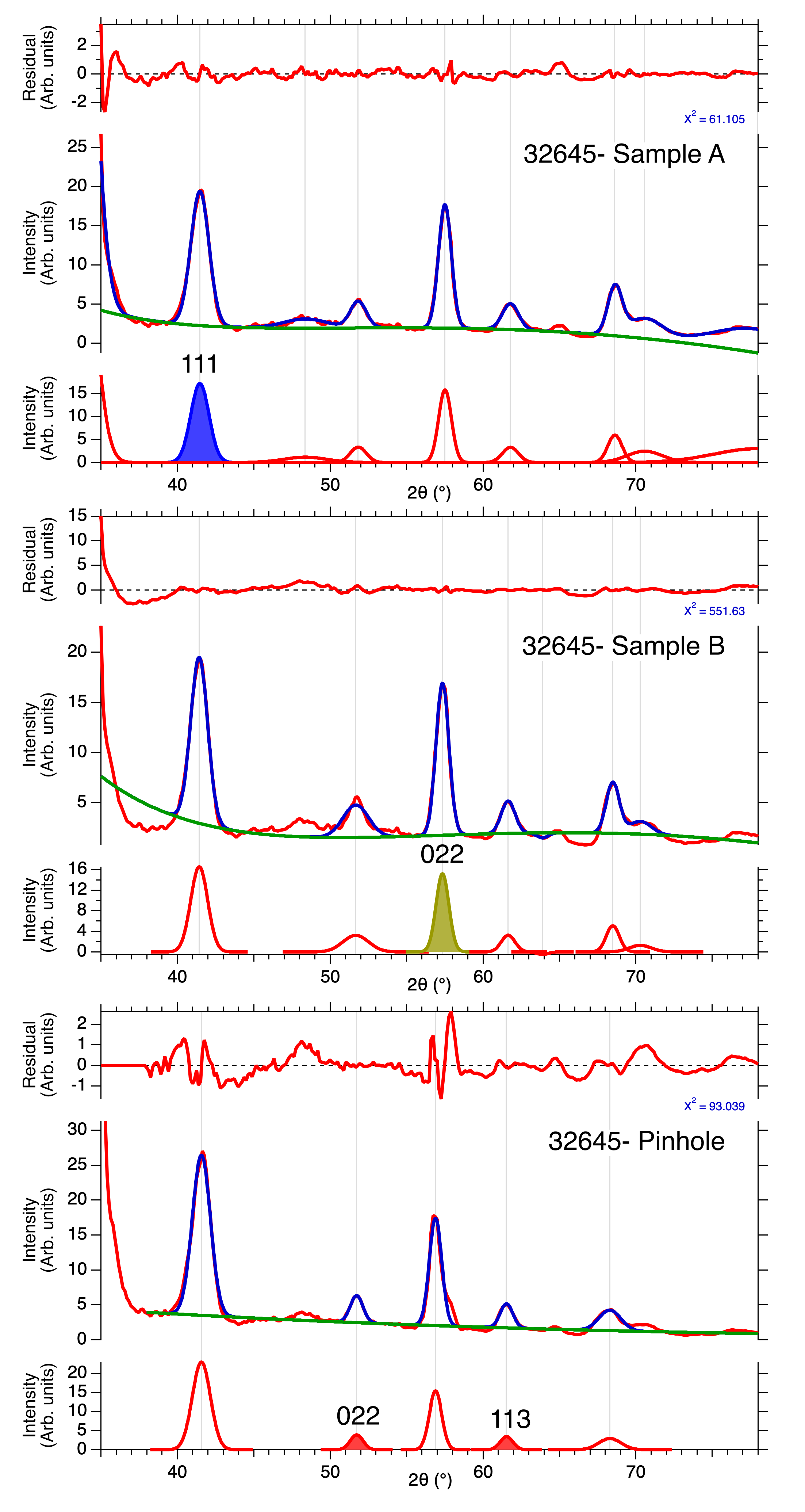}}   
  	\caption{\textbf{X-ray diffraction lineouts for 32645 and multipeak fits}. (\textbf{Top}) Shocked diamond projection (Sample A). (\textbf{Center}) Unshocked diamond projection (Sample B). (\textbf{Bottom}) Pinhole $2\theta-\phi$ projection: ambient platinum. For each graph, the measured lineout (red) is overlaid with the gaussian fit results (blue). The bottom plot shows the inferred individual peaks and highlights (color shaded fill) the peak(s) that are tracked and used in the post-processing analysis. For the Sample A (shocked diamond) we use the most intense peak which we assign to the 111 reflection of compressed diamond near 42$^{\circ}$. For sample B (unshocked diamond) we use the 022 line near 57$^{\circ}$ because the most intense 002 line overlaps with the 111 Pt line. For the pinhole we focus on the vicinity of the compressed sample 111 line and track the 022 and 113 lines at 51.71$^{\circ}$ and 61.51$^{\circ}$. Here the lineouts were averaged over a wide range of azimuth:  $\phi \in [-130^{\circ}, 130^{\circ}]$.  }\label{fig:XRD_MCD_OutPutMultipeakFits_30974}	 
   	
  \end{figure}

  The area, position and linewidth of relevant peaks of interest are determined with gaussian fits after removing a residual background, for example with a third- or fifth-order polynomial function of $log(2\theta)$ (see Fig.~\ref{fig:XRD_MCD_OutPutMultipeakFits_30974}).

  We describe in this section multiple approaches to determining the evolution of the XRD intensity for the shock compressed sample, including simple scaling to the brightness of the x-ray source, comparison with the signal from the platinum pinhole (see Figs.~\ref{fig:GeoSketchNotToScale} and \ref{fig:XRD_MCD_OutPut_32645}) and comparison between the signal of the shocked and unshocked fractions of the sample.
   \subsection{Scaling of the XRD intensity by the compressed sample volume }\label{sec:volumescaling}

   Many PXRDIP experiments\cite{rygg2012,coppari2022} utilize ramp compression with a thin sample layer so that the entire sample is at peak compression during the XRD snapshot. However, when using single or multi-shock compression with a thick sample, different compression states may be present during the XRD flash. 
   
   In the single shock compression experiment discussed here as an example, the XRD snapshot was collected about 2 ns prior to the shock breakout from the sample and a significant fraction of the sample is still at ambient. To enable meaningful comparison with other experiments for which the shock had traveled further through the sample when the XRD pattern was collected, it is necessary to scale the observed XRD intensity to the volume of the sample that is compressed to the conditions of interest.

    We scale the area under the peak of the shocked sample $A$ by dividing it by the effective thickness for the shocked fraction of the sample $d_S$ and the intensity of the ambient diamond peak by $d_{U}$ (see Section \ref{Sec:Compression} for details on $d_{S}$ and $d_{U}$).
   
  \subsection{Scaling of the XRD intensity by the x-ray source brightness}\label{sec:scaling of the XRD intensity to the expected number of x-ray photons}
  A simple approach to compare the XRD intensity of a line of interest between different laser experiments is to scale the XRD intensity to the x-ray source  brightness. This simple scaling of the XRD intensity to the XRS brightness may become unreliable if large shot-to-shot variations in the conversion efficiency of the laser energy into He-$\alpha$ x-rays occur. If possible, we recommend using an absolutely calibrated x-ray spectrometer to measure the x-ray source  brightness for each experiment. Alternatively, one can use  the measured laser energy ($E_{LX}$) delivered to the secondary target (x-ray source) to estimate the energy converted into He-$\alpha$ x-rays ($E_{XRS}$) using the conversion efficiency ($CE$) :
$
  	E_{XRS}=CE ~E_{LX}$.
    
 Previous measurements of the conversion efficiency for the optimized configuration at the Omega and EP laser facilities\cite{coppari2019} found that $CE\approx1\%$ for Cu and 0.6\% for Ge. As illustrated on Figs.~\ref{fig:GeoSketchNotToScale} and \ref{fig:XRD_MCD_OutPut_32645} computing the actual number of photons which contribute to the diffraction signal for any pixel detector is complex. However, we can estimate that about $10^{11}$ of x-ray photons are collimated by the pinhole having a solid angle of about $10^{-4}$ steradians when using 2.5 kJ of laser energy with a 0.6\% conversion efficiency which corresponds to about $10^{16}$ x-ray photons at 10.24 keV (distributed over $4\pi$). 

 \subsection{Evolution of the XRD intensity relative to the ambient pinhole signal}
 \label{sec:Evolution of the XRD intensity relative the ambient platinum signal}
 
 Another possibility to determine the evolution of the XRD intensity for the compressed sample across a series of experiments is to use the observed signal of the pinhole material as reference. If multiple pinhole lines are observed with similarly good signal-to-background ratios it may be beneficial to track the intensity of several lines. In that case it is necessary to determine the expected relative intensity of the multiple pinhole lines. This can be done by computing the theoretical pattern for the ambient pinhole material using tabulated atomic form factors. Neglecting preheat due to the sample drive of x-ray source ablation plasmas, we usually assume that the pinhole material is at ambient. 
    
It may also be useful to relate the expected intensity of the pinhole lines obtained with two different x-ray sources within a given dataset. To do this, we need to determine the expected relative intensity of the XRD signal from the pinhole at the energy of the x-ray sources and this requires evaluating the efficiency of scattering. As described in Sec.~\ref{Sec:Geometry and data format} and in Ref. \cite{rygg2020}, the XRD signal from the pinhole originates from multiple locations along the inner wall and rims. However, except near the boundaries of the $2\theta-\phi$ coverage the dominant contribution originates from the regions of the interior of the pinhole \textit{walls} having a direct line of sight to the source and the detector. It is therefore reasonable to approximate the configuration as similar to that of a diffractometer operating in reflection with an optically thick sample for which the signal is less attenuated when the sample is less absorbing with a intensity scaling as the attenuation length $1/\mu$ at $E_{XRS}$. For example, comparing the attenuation of Pt at the energy of Ge and Cu x-ray sources yields we find $1/\mu=2.63 \mu m$ with a Cu source and  $1/\mu=4.38 \mu m$ with a Ge source, so that the scattering efficiency contributes to a weaker diffraction of the Pt by a factor of $2.63/4.38$ when using the Cu source.
  
In addition, the energy dependence of the form factor may contribute to slight differences in the theoretical XRD cross-section per unit volume when using different x-ray sources. It is therefore recommended to compute the theoretical spectrum of the pinhole material at each x-ray source energy to infer the theoretical intensity ratio between the various peaks and accurately determine the scaling factors to be applied to the sample lines.

  \subsection{Evolution of the XRD intensity relative to an uncompressed reference layer: example with unshocked diamond}\label{sec:Evolution of the XRD intensity relative the unshocked diamond signal}
   \begin{figure}[!htb]
  	\centerline{\includegraphics[width=.49\textwidth,trim=0 0 0 0,clip]{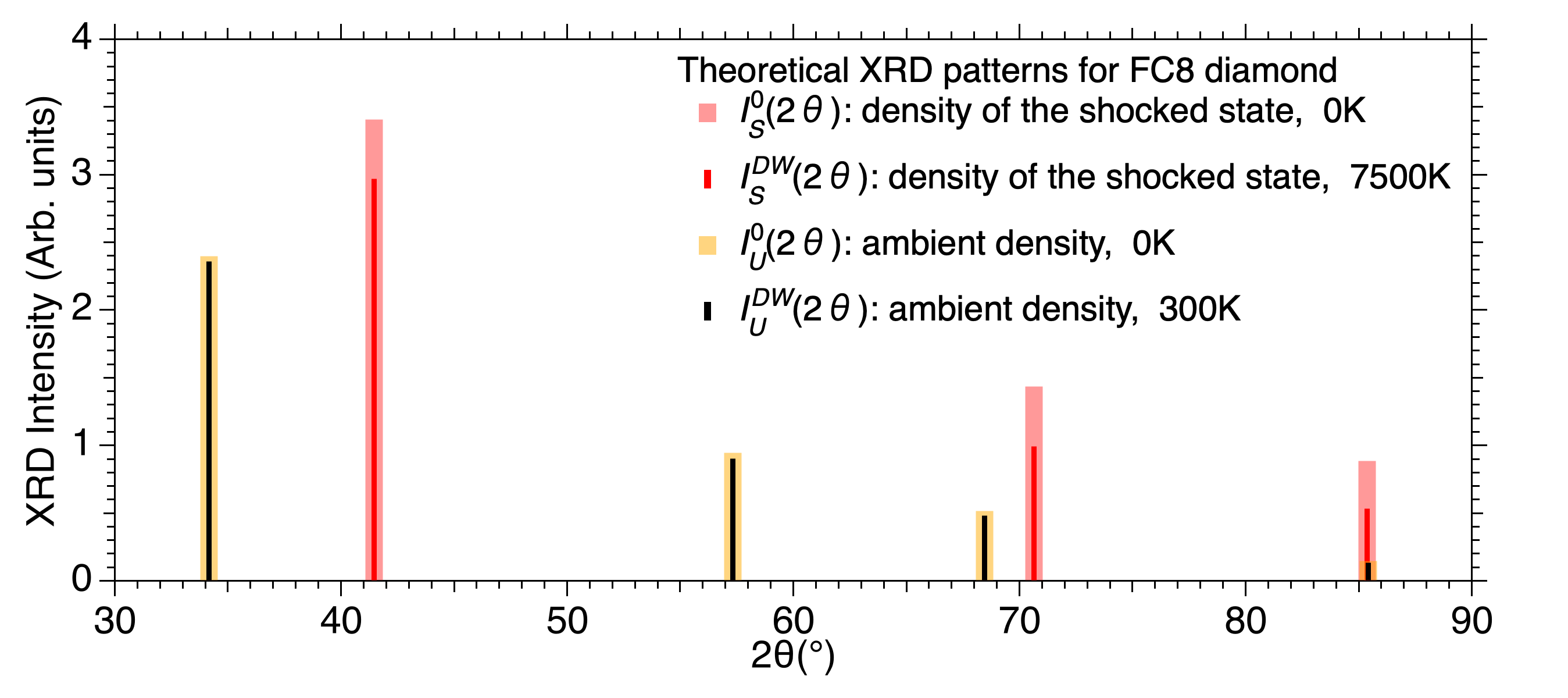}}     
  	\caption{\textbf{Simulated effect of thermal vibration on the XRD signal.} Comparison between the expected XRD patterns for the shocked and unshocked portions of the sample during the XRD snapshots $I_S^{DW}(2\theta)$ and $I_U^{DW}(2\theta)$ with the corresponding patterns at 0K $I_S^{0}(2\theta)$ and $I_U^{0}(2\theta)$. We use $T_{D0}=1860~ K$ as the Debye temperature at ambient and approximate\cite{benedict2014,yakub2017} the Debye temperature of the shock compressed state near $6.5~g/cm^3$ and 7500 K by $T_{D}=3000~ K$.  For the shocked sample the lattice parameter is determined by the observed diffraction angle for the 111 peak. For the unshocked sample we use $a=3.567~\rm\AA$. On this scale the ambient density patterns at 0 K and 300 K are indistinguishable. }\label{fig:DW_XRD_Pattern30974}	 
  \end{figure}

In contrast to ramp compression XRD experiments for which the entire sample is usually dynamically compressed when the XRD snapshot is collected, shock compression XRD data of polycrystalline samples often include a contribution from the sample that is still at ambient  if the XRD snapshot is captured before the shock has traveled through the entire sample. In the example dataset on shock compressed diamond the snapshot was collected when the shock had traveled about half-way through the sample (Fig.~\ref{fig:Hyades_32645}) and a significant contribution from the ambient diamond is clearly visible in the XRD data (Fig.~\ref{fig:XRD_MCD_OutPut_32645}). 

Using the signal of the uncompressed portion in a shock experiment as an \textit{in-situ} reference for the XRD intensity may enable a quantitative absolute determination of the XRD intensity of the shocked sample if (1) the relative volume of the shocked and unshocked fractions can be precisely determined by velocimetry or hydrodynamic simulations; and (2) the ambient sample is a good powder.  If the ambient sample is a single crystal or highly textured the XRD intensity of the ambient component will depend on the orientation of the target, and may vary from experiment to experiment.

In order to determine the intensity of the XRD signal for the shocked sample relative to the ambient diamond we need to account for both the higher density and higher temperature of the shocked sample relative to the undriven diamond ahead of the shock front. In the example dataset, the shock state is $\sim$ twofold compressed and the shock temperature is $\sim$ 7500K. 

As illustrated in Fig.~\ref{fig:DW_XRD_Pattern30974}, the damping of the XRD lines at elevated density and temperature can be described by a Debye-Waller approach. Using the squared mean interatomic displacement ($u^2_S(T)$) of the shocked sample we can compute the XRD pattern of the shocked sample at elevated density and temperature ($I_S^{DW}(2\theta)$) as:
\begin{equation}
	I_S^{DW}(2\theta)=I_S^{0}(2\theta)~e^{\left(-(4 \pi sin(2\theta/2)/\lambda)^2~	u^2_S(T)\right)}
\end{equation}
Similarly, using the squared mean interatomic displacement ($u^2_U(T_0)$) of the unshocked fraction at ambient temperature and density we can model the XRD pattern of the unshocked sample ($	I_U^{DW}(2\theta)$) as:
\begin{equation}
	I_U^{DW}(2\theta)=I_U^{0}(2\theta)~e^{\left(-(4 \pi sin(2\theta/2)/\lambda)^2~	u^2_U(T_0)\right)}
\end{equation}
where $I_S^{0}(2\theta)$ and $I_U^{0}(2\theta)$ are the XRD patterns calculated at 0 K, and $\lambda$ is the wavelength of the x-ray source. Here, we neglect thermal diffuse scattering\cite{wark2025,heighway2025}. We note that no noticeable thermal expansion is detected in the observed diffraction lines for the unshocked diamond.

We describe below a simplified model to account for both density and temperature effects. Limitations for this model include (1) a formalism adapted only to monoatomic cubic crystals; (2) neglecting thermal diffuse scattering\cite{wark2025,heighway2025} because the use of the 2D SNIP algorithm to remove the large ablation plasma background does not allow us to isolate signatures of thermal diffuse scattering; and (3) using a single temperature Debye-model even though describing material properties may require more advanced modeling to account for non-harmonic effects or more complex phonon-density of states (see Refs.\cite{benedict2014,yakub2017} for a description of advanced Debye models for compressed Diamond).

Multiple approaches can be used to evaluate the amplitude of the thermal motion of the ions and estimate the squared mean interatomic displacement for the Debye-Waller model, including theoretical modeling of the vibrational properties or interrogating explicit molecular dynamics simulations. Here we describe a simplified theoretical approach, valid for monoatomic cubic crystals adapted from Eqs. 3.21 and 11.77 in Ref.\cite{warren1969} .
The squared mean interatomic displacement ($u^2_S(T)$) of the shocked sample at elevated density and temperature is:
  \begin{align}
  	u^2_S(T)&=3 \hbar^2  T / (m_{at}  k_B   T_D^2) ~ \left(\Phi(x)+x/4\right)
  \end{align}
where $m_{at} =M/N_A$ is the mass per atom, $T$ is the temperature of the shocked sample (in our example dataset the temperature of the shocked sample is near 7500 K), $T_{D}$ is the Debye temperature at high density, $x=T_D/T$ and the function  $\Phi(x)$ is defined as:
  \begin{equation}
  	\Phi(x)=\frac{1}{x}\int_0^x \frac{z } {(e^z-1)}dz
  \end{equation} 
    
Using this formalism the squared mean interatomic displacement ($u^2_U(T_0)$) of the unshocked fraction can be computed by setting the temperature to $T_0=300$ K and using the Debye temperature at ambient density $T_{D0}$ and $x_0=T_{D0}/T_0$:
  \begin{align}
  	u^2_U(T_0)&=3 \hbar^2  T_0 / (m_{at}   k_B   T_{D0}^2) ~ (\Phi(x_0)+x_0/4)
  \end{align}
  
The effect of the increased density of the compressed state is to increase the Debye temperature which reduces the amplitude of the squared mean interatomic displacement. For example, while diamond's Debye temperature\cite{benedict2014,dewaele2008a} at ambient is $T_{D0}=1860 K$, it is much higher near two fold compression. Combining previous models to obtain a single Debye temperature for the shock compressed state near $6.5~g/cm^3$ and 7500 K we find\cite{benedict2014,yakub2017} $T_{D}=3000\pm500 K$.

  Altogether, it may be convenient to define a scaling factor ($R_{ij}^{DW}$) between the expected intensity per unit volume of the $\rm i^{th}$ line for the compressed state relative to the $\rm j^{th}$  line of the ambient material ahead of the shock. In our example we define the ratio between the expected intensity of the 022 line of unshocked diamond and the 111 line of the shocked diamond
  \begin{equation}
  	R_{12}^{DW}=\frac{I_U^{DW}(2\theta_U^{022})}{I_S^{DW}(2\theta_S^{111})}
  \end{equation} 
  Here, $2\theta_S^{111}$ is the diffraction angle of the first peak of the shocked sample pattern while $2\theta_U^{022}$ is the diffraction angle of the second peak of the unshocked pattern.
 Combining $R_{ij}^{DW}$ and the thicknesses of the compressed and ambient fractions of the sample, we can determine the XRD intensity of the observed compressed peak of the shocked sample in an absolute way. The solid fraction ($SF$), \textit{i.e.} the fraction of the shock compressed solid that is in a crystalline solid state is 
  \begin{equation}
 	SF_{ij} =\frac{A^i/d_S}{A^j_{U}/(d_U~R_{ij}^{DW})}
 \end{equation}
where $A^i$ is the measured area of the $\rm i^{th}$ shocked sample line, $A^j_{U}$ is the measured area of the unshocked sample $\rm j^{th}$ line,  $d_S$ is effective thickness for the shocked fraction of the sample and $d_U$ is the thickness of the unshocked fraction of the sample (see Fig.~\ref{fig:Hyades_32645}).

  In the example dataset discussed above we find $SF_{12} = 0.6\pm0.05$ while theoretical models at these conditions assuming a hydrostatic Hugoniot state, neglecting strength and kinetic effects suggest that the solid fraction should be closer to 0.9. See Ref.\cite{millot2025} for an application of the methodology detailed here to shock compressed diamond up to 1 TPa. The discrepancy between the experimental estimate and the theory illustrates the difficulty of determining absolute PXRD intensities at kJ and MJ laser facilities. 
  
  We consider several caveats for the current analysis framework. First, our experimental configuration and data processing approach (including the use of the 2D SNIP algorithm) may favor preferentially the \textit{spotty} XRD signal from the unshocked portion of the diamond sample compared to the broader signal from the shocked sample. Second, our approach ignores any preferred orientation effects which may be present (even though we verified that the observed intensity ratio between the  111 and 022 reflections for the ambient diamond is consistent with that of a perfect powder when averaged over a large $\phi$ range). Further methodology development are needed to address this limitation because carefully considering crystallographic texture is key when analyzing powder XRD for shock compressed materials because the shock induced response in the material may significantly modify its texture\cite{wehrenberg2017a,heighway2022}. Third, because our experimental configuration generates high levels of highly structure background and the use of the SNIP algorithm we cannot quantify and therefore neglect any contribution from the thermal diffuse scattering. Finally, using a single scalar -- the \textit{Debye Temperature $T_D$} -- to describe the temperature dependence of lattice vibrations may be over-simplistic. Improved modeling and data analysis methodology to address these caveats is ongoing but falls outside the scope of this work.
  
     \section{Conclusion}
     
In this work we describe data analysis methods for single-frame, powder x-ray diffraction in transmission geometry of dynamically compressed samples at the Omega and NIF laser facilities, with a focus on improving the fidelity of the diffraction intensity determination. 

The methods discussed should be applicable to experiments at the Laser Fusion Research Center (LFRC) \cite{sun2024} given the strong similarities between the LFRC and NIF experimental approaches. They should also prove valuable--with some modifications due to different choices for the position and collimation of the x-ray source--to analyze experiments at the LULI2000\cite{denoeud2021} and ORION\cite{foster2021} laser facilities, and for multi-frame, time-resolved XRD at the Omega\cite{polsin2025} and NIF\cite{vennari2024}. Future developments may include diagnosing and quantifying how crystallographic texture evolves under dynamic compression.

 \paragraph*{Acknowledgments}
We acknowledge insightful discussions with J. Ryan Rygg and the TARDIS team at LLNL. This work was prepared by LLNL under Contract No. DE-AC52-07NA27344. We used our LLNL AnalyzeVISAR and AnalyzePXRDIP codes for data analysis. The authors have no conflicts to disclose. Raw data were generated at the OMEGA Laser Facility. Derived data supporting the findings of this study are available from the corresponding author upon request and with permission from the OMEGA Laser Facility.
 
\bibliography{Zotero2025.bib}

\end{document}